\documentclass[reprint,nofootinbib,floatfix,aps,prd]{revtex4-2}

\usepackage{graphicx}
\usepackage{amsmath,amssymb,mathtools}

\usepackage[caption=false]{subfig}

\usepackage{booktabs}
\usepackage{dcolumn}
\usepackage{multirow}
\usepackage{array}
\usepackage{adjustbox}
\usepackage{rotating}

\usepackage{orcidlink}
\usepackage[T1]{fontenc}

\newcommand{\lcdm}{\texorpdfstring{$\Lambda$CDM}{LCDM}}

\newcommand{\wowacdm}{\texorpdfstring{$w_0w_a$CDM}{w0waCDM}}

\newcommand{\HzeroDN}{H_0^{\rm DN}}

\newcommand{\kmsMpc}{\,\mathrm{km\,s^{-1}Mpc^{-1}}}

\begin{document}

\preprint{000-000-000}

\title{The DESI results impact the local determination of $H_0$}

\author{Michael S. Turner}
\email{mturner@uchicago.edu}
\affiliation{Kavli Institute for Cosmological Physics,
The University of Chicago, Chicago, IL  60637-1433}
\affiliation{Department of Physics and Astronomy, University of California, Los Angeles, Los Angeles, CA  90095-1547}

\author{Dragan Huterer}
\email{huterer@umich.edu}
\affiliation{Department of Physics and Leinweber Institute for Theoretical Physics,
  University of Michigan, 450 Church St, Ann Arbor, MI 48109, USA}

\date{\today}

\begin{abstract}
Measurements of baryon acoustic oscillations (BAO) by the Dark Energy Spectroscopic Instrument (DESI)
have revealed evidence for dark energy that evolves.  If local distance measurements are analyzed with the $w_0w_a$ models preferred by the DESI measurements, the value for the Hubble constant can be as much as $2.5\kmsMpc$ smaller than the value obtained assuming \lcdm. When these $w_0w_a$ models are further constrained by cosmic microwave background (CMB) and type Ia supernova (SNIa) data, the downward shift is 
$1.1 \pm 0.38\kmsMpc$ (DESI + CMB) and $0.5 \pm 0.1\kmsMpc$ (DESI + CMB + SNIa). The dependence of local determinations of $H_0$ on the background cosmology, combined with the fact that the low-redshift cosmology is not well constrained, is relevant to the Hubble tension.
\end{abstract}

\maketitle

\section{Introduction}
\label{sec:intro}

The Hubble constant $H_0$ is arguably the most important number in cosmology, setting ages and distances in the Universe.  The almost hundred years of measuring the local Hubble flow ($z \le 0.1$) has seen determinations of $H_0$ go from the near qualitative to precision measurements based upon a carefully calibrated local distance scale \cite{Freedman_Science}.  The precision of some direct measurements of $H_0$ is now approaching 1\% \cite{H0DN:2025lyy}.  

Another approach, that based upon cosmic microwave background (CMB) and baryon acoustic oscillation (BAO) measurements has yielded similar or better precision determinations of $H_0$.  The CMB method, in effect uses a precise determination of the expansion rate around the time of last scattering ($z_{LS} \simeq 1090$) and the \lcdm\ cosmology to extrapolate the expansion rate to the present epoch.  

The discrepancy between the indirect approach, which finds $H_0 \simeq (68 \pm 0.5)\kmsMpc$ \cite{Planck:2018,DR2,AtacamaCosmologyTelescope:2025blo}, and the local, distance-ladder type approaches which find
$H_0=(73.50\pm 0.81)\kmsMpc$ \cite{H0DN:2025lyy}, is known as the Hubble tension \cite{Verde:2019ivm,DiValentino:2021izs,Hubble_tension}.  It is one of the most pressing issues in cosmology today. The resolution could be due to systematic errors in either (or both) the direct and indirect measurements, or an indication that something is missing in the $\Lambda$CDM model \cite{PNAS_Turner}.  Of course, viewed with a historical perspective, where determinations of $H_0$ has evolved by almost a factor of ten \cite{Freedman:2010xv}, the 7\% present agreement of the two very different methods is a powerful end-to-end test of the current cosmological paradigm $\Lambda$CDM and testimony to the fact that we are indeed in the era of precision cosmology.

Because the bar in cosmology is now higher, a host of modifications to the $\Lambda$CDM cosmology around last scattering and earlier, e.g., an early dark energy \cite{Hubble_tension}, have been proposed to increase the CMB value for $H_0$  to match the local determinations.  Our focus here is on low redshifts and the value of $H_0$ determined from distance measurements at $z\lesssim 0.1$.  

\section{$H_0$ and the background cosmology}
\label{sec:background}

To address the dependence of these direct determinations of $H_0$ upon the background cosmological model, we use the Hubble Distance Network (H0DN) combination of low-redshift measurements \cite{H0DN:2025lyy}, and an accompanying computer code that uses these measurements and their covariances in order to extract the Hubble 
constant.\footnote{\url{https://github.com/StefCas789/H0DN}}. Our results should apply more broadly to local determinations of $H_0$ based upon other datasets, e.g., those in Refs.~\cite{Riess:2021jrx,Freedman:2024eph}.

\begin{figure}[t]
\includegraphics[width=0.5\textwidth]{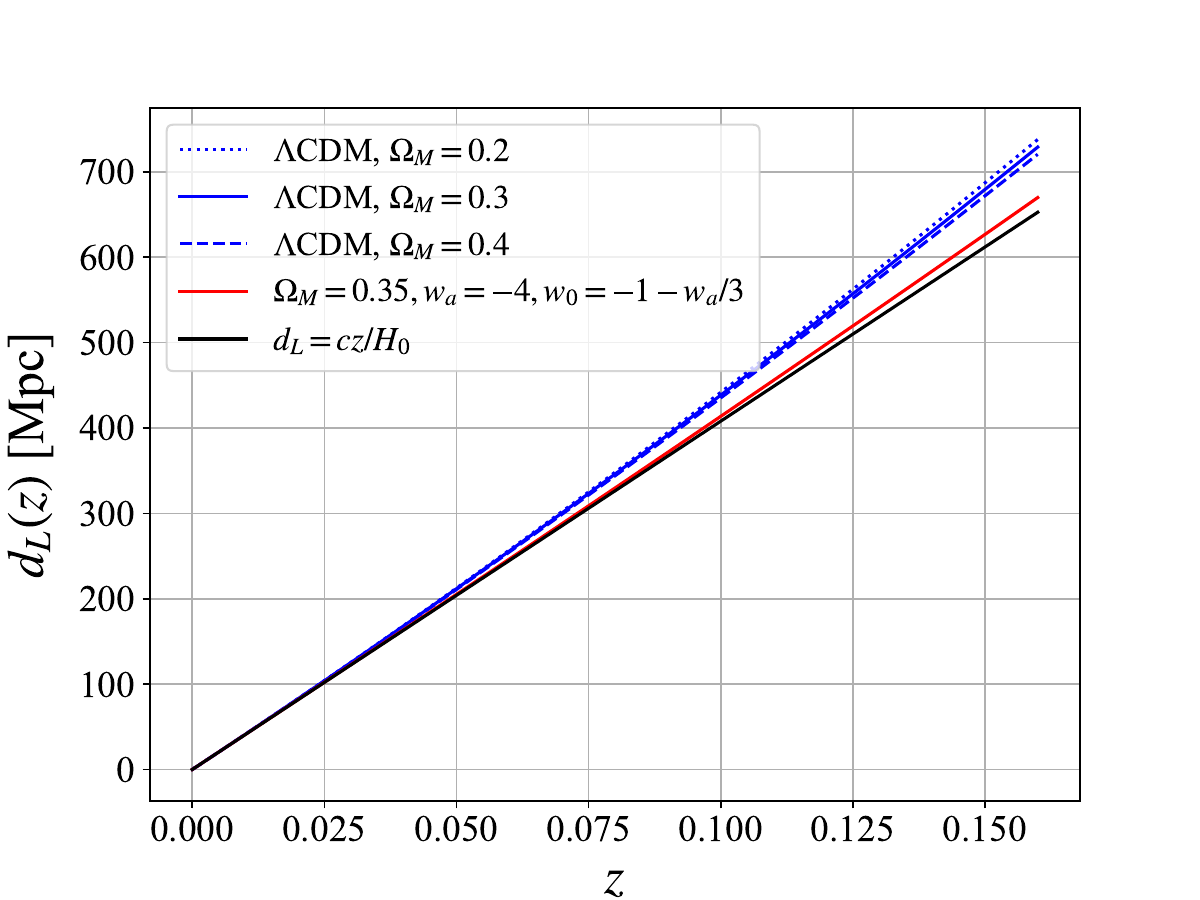}
\caption{Luminosity distance vs. redshift for:  \lcdm\ with $\Omega_M = 0.2, 0.3, 0.4$, a \wowacdm\ model, and for reference, the Hubble Law.  Note the tiny dependence in $\Lambda$CDM upon $\Omega_M$.}
\label{Hubble_diagram}     
\end{figure}

The H0DN data span redshifts $z \simeq 0-0.1$. Thus, one would expect that the Hubble constant determined from this data to depend little upon the background cosmology, since the cosmology-dependent luminosity distance, $d_L(z) = (1+z) \int_0^z dx/H(x)$, differs from the Hubble Law, $d_L(z) = cH_0^{-1}z$, by $\mathcal{O}(z^2)$ which at $z = 0.05$ is only a 4\% difference, cf.~Fig.~\ref{Hubble_diagram}. Because the precision of direct determinations of $H_0$ is now approaching 1\%, this expectation is worth testing and quantifying. 

Further, as illustrated by the recent combined constraints from BAO, CMB and type Ia supernovae (SNIa) (e.g.\ \cite{DESI:2025fii}), the background cosmology is not well constrained at redshifts less $z \simeq 0.5$.  We believe this to be the case for two reasons:  First, the most precise techniques, e.g., BAO and SNIa, require many objects and a large volume, which become impossible at low redshift, and second, cosmological leverage decreases with redshift since the deviations from the Hubble Law become smaller and harder to measure.

Here and throughout, we use \lcdm\ with $\Omega_M = 0.3$ as our fiducial cosmology and the H0DN data to illustrate how local measures of $H_0$ depend upon the background model.  ($\Omega_M$ is the ratio of the matter density to the cricial density.) For this model and the H0DN data, $H_0 = (73.50\pm 0.81)\kmsMpc$.\footnote{In Ref.~\cite{H0DN:2025lyy}, the same data are analyzed using the $(q_0 = -0.55, j_0 =1)$ representation of \lcdm\ for $\Omega_M = 0.30$, where $q_0$ and $j_0$ are respectively the deceleration and jerk parameter evaluated today.  Their result, $H_0 = 73.50\kmsMpc$, is essentially identical to our result which uses the exact expression for $d_L(z)$.}  This is our baseline value for $H_0$.

The consensus cosmology, \lcdm, has one parameter in addition to $H_0$, the matter density $\Omega_M$. Analyzing the same H0DN data, but now varying $\Omega_M$ over the generous range $\Omega_M =  0.2 - 0.4$, the value obtained for $H_0$ varies from $73.79\kmsMpc$ to $73.21\kmsMpc$, $\Delta H_0 = \pm 0.29\kmsMpc$ or about $0.3\sigma$; see Fig.~\ref{Hubble_diagram}.  Not only is the $\Omega_M$--dependence small, but also $\Omega_M$ is well determined by other cosmological data.

Beyond $\Lambda$CDM, there is the flat $w$CDM model, where $w$ is the equation-of-state of dark energy.  Here, the dependence upon the equation-of-state parameter is more significant.  Assuming $w$CDM is the true cosmology and using the same H0DN data, as $w$ varies from $-1$ to $+0.4$ the derived value of $H_0$ decreases linearly with increasing $w$ from $73.50\kmsMpc$ ($w = -1$) to $70.84\kmsMpc$ ($w = 0.4$), a downward shift of upto $2.7\kmsMpc$ from the baseline \lcdm\ value. This is a significant dependence, but in the $w$CDM cosmology, high-redshift SNIa data constrain $w$ to be very close to $-1$ ($\pm 0.1$ or less) \cite{Brout:2022vxf,DES:2024jxu}, and any shift in $H_0$ is very small for the allowed range, $\pm 0.2\kmsMpc$ or less.

A third parameter, $w_a$, is often added to test for the possible variation of dark energy with time or an equation-of-state different from $w = -1$ \cite{Linder:2002et,Chevallier:2000qy}, $w(a) = w_0 + w_a (1-a) = w_0 + w_az/(1+z)$, where $a$ is the cosmic scale, normalized to unity today and redshift $z = a^{-1} -1$.  The \wowacdm\ cosmology is characterized by three parameters, $w_0, w_a$ and $\Omega_M$ in addition to $H_0$.  Relevant to the discussion above, because of the evolution of the equation-of-state, the value of $w$ today ($\equiv w_0$) can be significantly different from $-1$.

\begin{figure} [t]
\center\includegraphics[width = 0.5\textwidth]{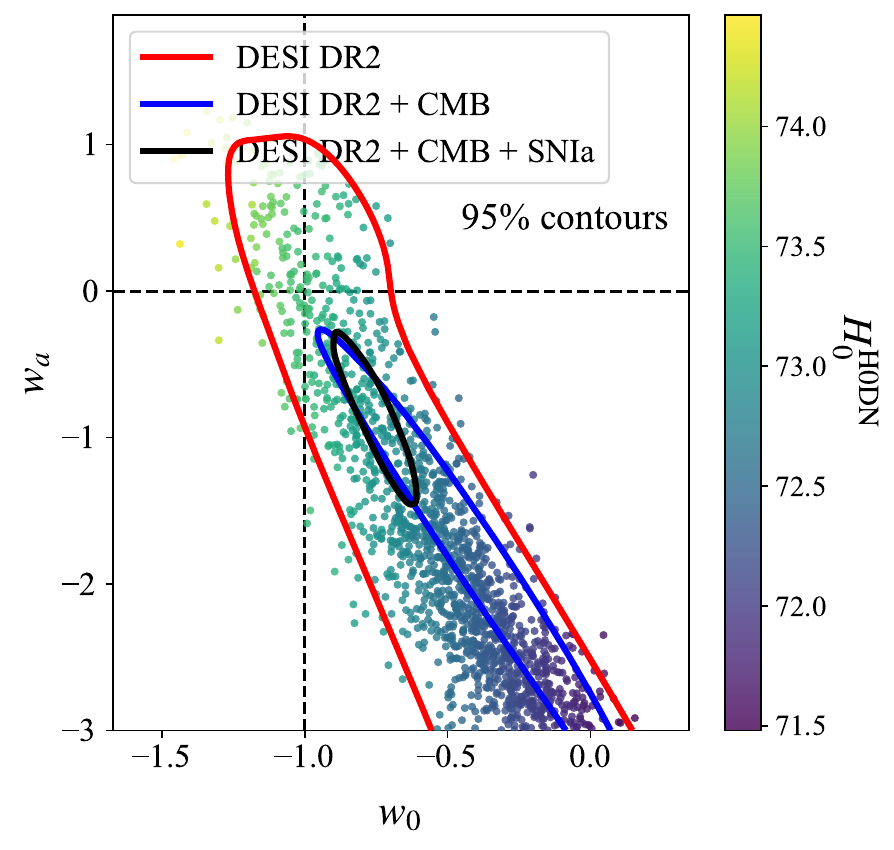}
\caption{95\% credible regions in the $w_0$--$w_a$ plane based upon the DESI DR2 BAO alone, with the CMB added, and with the DES SNIa added. Points with the corresponding color bar denote values of the Hubble constant in the H0DN analysis for a selection of models sampled from the DESI DR2 chain.  Note: $\Lambda$ corresponds to $w_0=-1$ and $w_a=0$.}
\label{DESI_ellipse}     
\end{figure}

The recent BAO measurements from the Dark Energy Spectroscopic Instrument's Data Release 2 (henceforth just DESI) \cite{DR2} prefer a contour in the $w_0$--$w_a$ plane whose degeneracy axis is $w_0 = -1 - w_a/3$ for $w_a<0$, which provides evidence for evolving dark energy; see Fig.~\ref{DESI_ellipse}.  Considering the DESI results alone, the contour does not close for values of $w_a < -3$; adding a CMB constraint \cite{Planck:2018,AtacamaCosmologyTelescope:2025blo} closes the contour around $w_a = -3$; and further, adding the type Ia supernova data from the Dark Energy Survey (DES SNIa; \cite{DES:2024jxu}) results shrink it even further, cf.~Fig.~11 in Ref.~\cite{DR2}.

Fig.~\ref{DESI_ellipse} also shows the values of the Hubble constant extracted using the Distance Network code for samples from the DESI DR2 Markov chains, assuming DESI DR2 BAO data alone. The downward shift in the inferred value of $H_0$ from that when $\Lambda$CDM is assumed can be as large as $2.5\kmsMpc$ or almost three times the error reported by H0DN analysis.   A shift of this amount would clearly be of relevance to the Hubble tension.  

\begin{figure}[t]
\center\includegraphics[width = 0.5\textwidth]{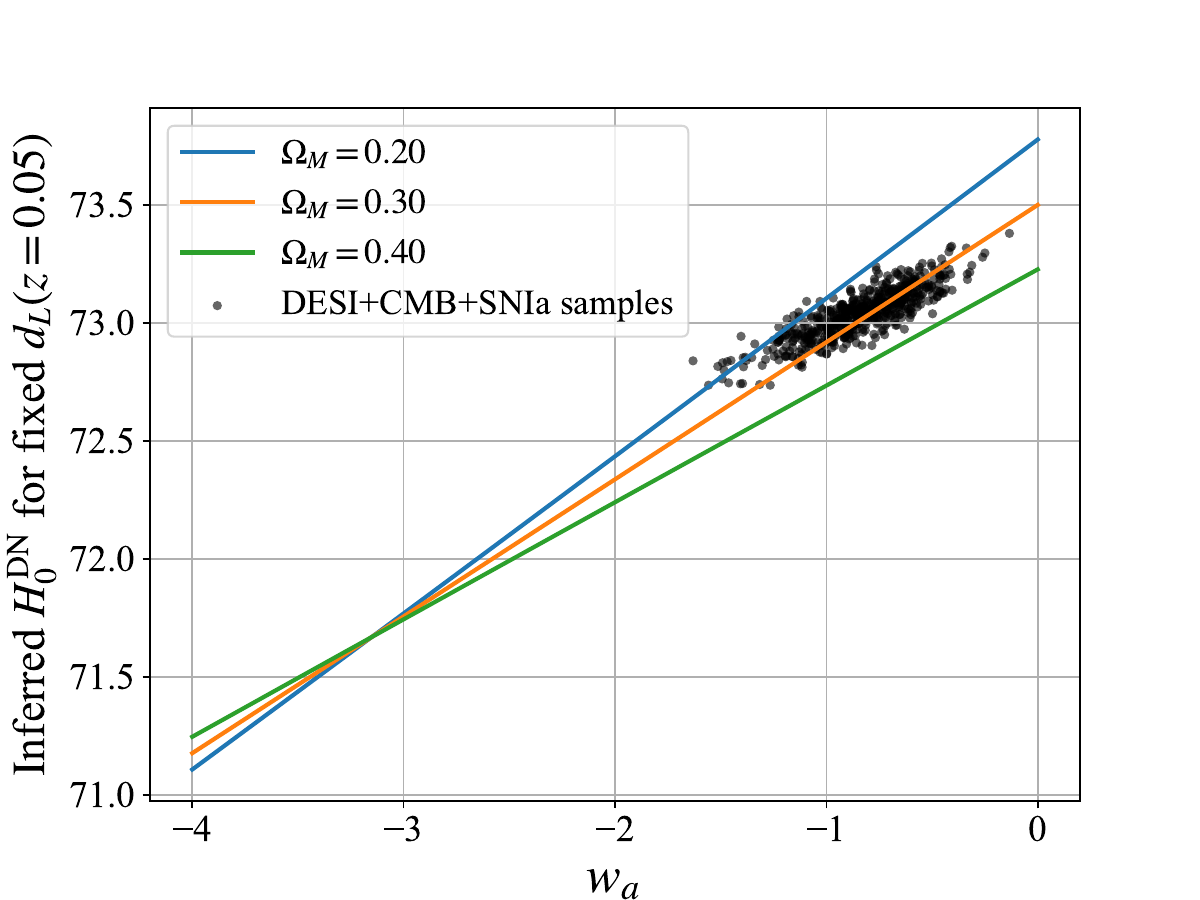}
\caption{ The lines show the dependence of $H_0$ upon $w_a$ when $w_0 = -1-w_a/3$ for $\Omega_M = 0.2, 0.3, 0.4$.  The black dots represent the derived Hubble constant from the Hubble Distance Network analysis for a sampling of models (i.e., $\Omega_M, w_a, w_0$) from the DESI+CMB+SNIa chain; note, only the value of $w_a$ is shown and $w_0$ is not constrained to be $-1-w_a/3$. When the H0DN data are analyzed assuming $\Lambda$CDM and $\Omega_M = 0.3$, $H_0 = 73.50\kmsMpc$.} 
\label{Pot_o_Gold}     
\end{figure}

Fig.~\ref{Pot_o_Gold} displays this trend by showing the value of $\HzeroDN$ vs.\ $w_a$, with $w_0=-1-w_a/3$ (i.e., along the direction along the DESI contour) and fixed $\Omega_M$. Also shown in Fig.~\ref{Pot_o_Gold} are a sample of models (i.e., $w_0,w_a, \Omega_M$) from the combined Markov chain constrained by DES+CMB+SNIa data.

The additional constraints from the CMB and SNIa significantly narrow the favored regions in the $w_0w_a$ plane,  cf.~Fig.~\ref{DESI_ellipse}.  Weighting the values of the Hubble constant in the Distance Network (DN) analysis with the cosmological parameters $(\Omega_M, w_0, w_a)$, we find for the DESI  + CMB + SNIa chains, the DESI + CMB chains, and for the DESI data alone,\footnote{Because the DR2 alone ellipse does not close, the prior $w_a \ge -3$ was imposed, limits the shift in $\HzeroDN$} respectively: 
\begin{equation}
\begin{aligned}
H_0^{\rm DN/DESI+CMB+SNIa} &= 73.03\pm 0.81 \pm 0.11,\\
H_0^{\rm DN/DESI+CMB} &= 72.44\pm 0.81 \pm 0.38,\\
H_0^{\rm DN/DESI} &= 72.64\pm 0.81 \pm 0.54,
\end{aligned}
\end{equation}
where all the values are in units of $\kmsMpc$.
Here the first error in each line comes from the Distance Network analysis \cite{H0DN:2025lyy}, while the second error comes from the uncertainty in the cosmological-parameter values for the cosmological probes. 

The central value for the local Hubble constant informed by the \wowacdm\ models is therefore lower than that for the fiducial \lcdm\ model used in the H0DN analysis, by from $0.5\kmsMpc$ to $1.1\kmsMpc$, depending upon which cosmological data are used to constrain $(\Omega_M, w_0,w_a)$.  There is also an additional statistical error due to the uncertainty in the cosmological model.

The DESI results call attention to the fact that something interesting may be going on with dark energy recently.  Not only is this important in trying to understand dark energy, but our results show that changes to the background cosmology at low redshifts could significantly impact local determinations of the Hubble constant.

\begin{figure}[t]
\includegraphics[width = 0.5\textwidth]{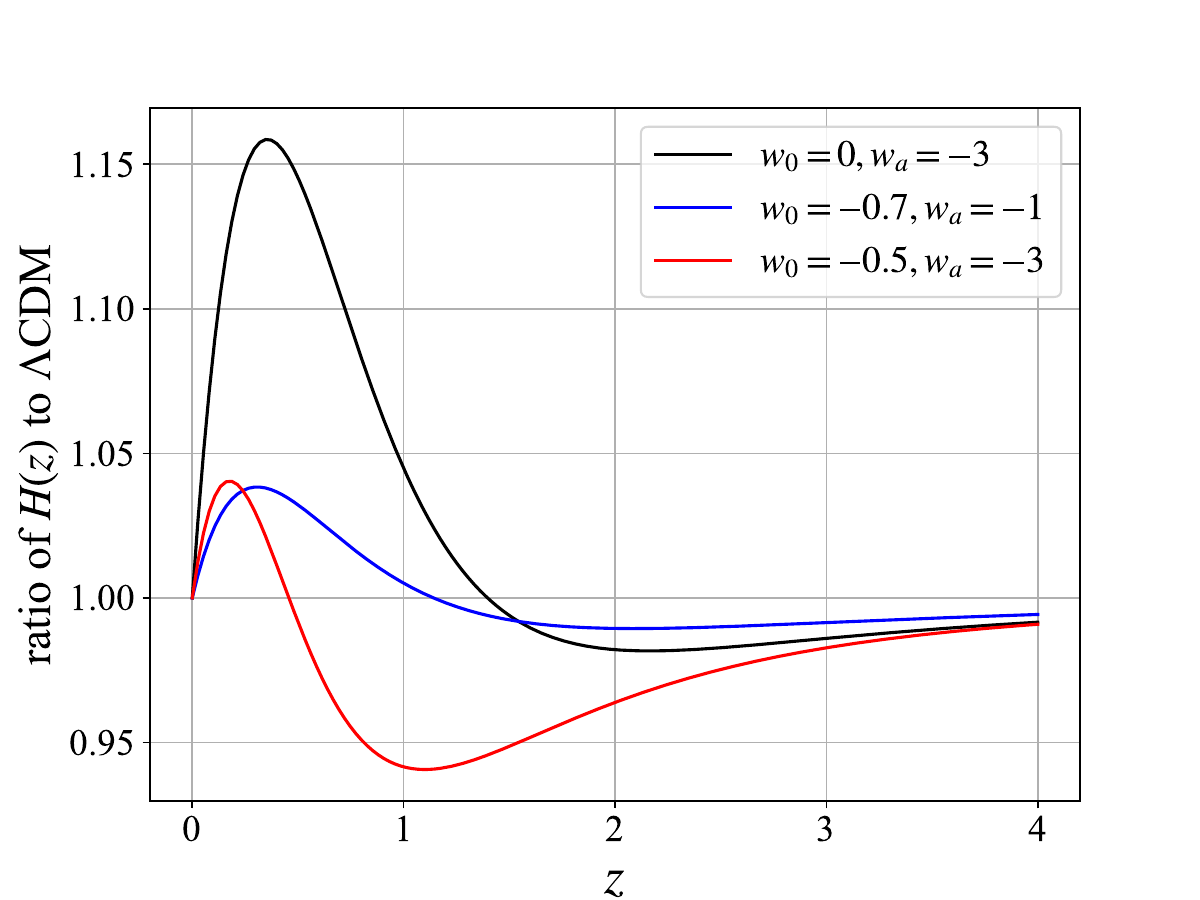}
\caption{Expansion rate relative to $\Lambda$CDM for three $w_0w_a$ models with $\Omega_M = 0.3$. For all three, the bump in the expansion rate relative to $\Lambda$CDM arises due to the dark energy density peaking at $1+ z_{\rm peak} =  w_a/(1+w_0+w_a)$, and as $w_a$ decreases, the peak's height increases and its width decreases.  The downward shifts in $\HzeroDN$ are: $1.8$ ($w_a=-3, w_0 =0$), $0.55$ ($w_a = -1, w_0 = - 0.7$), and $1.0$ ($w_a =-3, w_0=-0.5$), in $\kmsMpc$.}
\label{DESI_bump}     
\end{figure}

The DESI-preferred \wowacdm\ models feature a dark energy component that is peaked around redshift $z\simeq 0.5$, with a similar feature in the expansion rate.  To illustrate their effect on the recent cosmological history, Fig.~\ref{DESI_bump} shows the expansion rate for three $w_0w_a$ models consistent with the DESI results.

Thinking more broadly about changes in the cosmology at low redshift, consider dark energy that monotonically decreases, as might be the case with a thawing scalar field.  Such a possibility is both physically motivated and well studied. 
  
Scalar-field dark energy can be represented as a \wowacdm\ model, though not perfectly, by $\alpha \equiv 1 + w_0 + w_a = 0$ and  varying $w_0$  upward from $-1$ \cite{AbreuTurner}.  For $w_0 = -0.7, -0.6, -0.5$, the downward shifts in $\HzeroDN$ are $0.55, 0.73,$ and $ 0.90$ respectively in $\kmsMpc,$  for $\Omega_M = 0.3$. 

\section{Hubble tension}
\label{sec:disc}

Our primary result is that changes to the cosmological background model at low redshift can significantly impact the locally determined value of the Hubble constant, as we have illustrated with the DESI DR2 BAO results. Whether or not the DESI results hold up, this possibility raises an additional uncertainty in the direct measurements of the Hubble constant, and of course, is relevant to the Hubble tension.

One might be tempted to consider the 
\wowacdm\ models consistent with DESI data
as an alternative cosmology, and ask how they change the other side of the Hubble tension -- the CMB value of $H_0$.  There is peril in doing so.  Because of its unphysical behavior in much of the $w_0$--$w_a$ plane preferred by DESI \cite{AbreuTurner}, the \wowacdm\ models are not meant to describe the Universe over a large range of redshifts. Rather, they parameterize the dark energy equation-of-state at redshifts less than a few to probe for deviations from $\Lambda$ \cite{Chevallier:2000qy,Linder:2002et}. With that as a caution, we make a few remarks to illustrate how the two ends of the Hubble tension are connected.

The CMB value of $H_0$ is determined by the very precise Planck measurement of $\theta_*$ (to $\pm 0.03\%$), which is the ratio of the sound horizon to the comoving distance to the last scattering surface: $\theta_* \equiv r_s/r(z_{LS})$.  The sound horizon,  $r_s = \int_{z_{LS}}^\infty {c_s (z) dz/H(z)},$ depends upon ``early cosmology'' ($z_{LS} \simeq 1090$, $c_s(z)$ is the sound speed).  On the other hand, the distance to the last scattering surface, $r(z_{LS}) = \int_0^{z_{LS} }{dz/ H(z)}, $
depends upon the recent cosmological epoch of dark matter and dark energy -- and hence the DESI results.

Fixing $\Omega_M$ and assuming that there are no early modifications to the $\Lambda$CDM cosmology, moving along the degeneracy direction of the DESI ellipse for $w_0$ and $w_a$, the CMB value for $H_0$ decreases as $(1-w_a)^{-0.044}$,\footnote{Because $r(z_{LS})$ depends upon the integral of $1/H(z)$, a modification that involves a bump in the low-redshift expansion rate decreases $r(z_{LS})$, which leads to a decrease in the CMB value for $H_0$.} also cf. Table V in Ref.~\cite{DR2}.  This decrease in the CMB value for $H_0$ tends to maintain the Hubble tension.

The \wowacdm\ models allowed by DESI data illustrate that changes to the cosmology at low redshift can also impact the CMB determination of $H_0$.

\section{Concluding remarks}
\label{sec:concl}

The Hubble tension has been with us for more than a decade, and is rightfully continuing to garner attention.  It is well appreciated that modifications to the cosmological model at early times can change the CMB value of $H_0$ and impact the Hubble tension, perhaps even resolving it.  These ``early solutions'' largely address the tension by modifying the sound horizon with new cosmological physics before last scattering.  

However, the low-redshift cosmology is not well constrained, and we have shown that modifications to it consistent with the DESI results can impact measurements of the local value of the Hubble parameter significantly, and are also relevant to the Hubble tension conversation.  

Resolving the Hubble tension with physics beyond \lcdm\ may involve new physics at both high redshift and low redshift as well as the interplay between them.

\begin{acknowledgments}
DH acknowledges support from Department of Energy under contract DE-SC009193, and the Leinweber Center for Theoretical Physics at the University of Michigan.  We thank Tommaso Treu, Adam Riess and Wendy Freedman for their thoughtful comments, and Prakhar Bansal for useful conversations and assistance with setting up the MCMC runs.
\end{acknowledgments}

\bibliographystyle{apsrev4-2}
\bibliography{refs}

\end{document}